\documentclass[letter]{aa}
\usepackage{graphicx}
\usepackage{multicol}
\usepackage[varg]{txfonts}

\def\lae{\mathrel{<\kern-1.0em\lower0.9ex\hbox{$\sim$}}}
\def\gae{\mathrel{>\kern-1.0em\lower0.9ex\hbox{$\sim$}}}

\usepackage{amstext}

\begin{document}

   \title{The Close AGN Reference Survey (CARS)}

        \subtitle{Mrk 1018 halts dimming and experiences strong short-term variability\thanks{Based on observations collected at the European Organisation for Astronomical Research in the Southern Hemisphere under ESO programme(s) 098.B-0672 and 099.B-0159. The scientific results reported in this article are based on observations made by the \textit{Chandra X-ray Observatory} and the NASA/ESA \textit{Hubble Space Telescope.}}}
         
   \author{M. Krumpe  
          \inst{1}
          \and
          B. Husemann\inst{2}
          \and 
          G. R. Tremblay\inst{3,4}
          \and
          T. Urrutia\inst{1}
           \and 
          M. Powell\inst{4}
          \and 
          T. A. Davis\inst{5}
          \and 
          J. Scharw\"achter\inst{6}
          \and 
          J. Dexter\inst{7}
          \and 
          G. Busch\inst{8}
          \and 
          F. Combes\inst{9}
          \and 
          S. M. Croom\inst{10,11}
          \and 
          A. Eckart\inst{8,12}
          \and 
          R. E. McElroy\inst{10,11}
          \and 
          M. Perez-Torres\inst{13}
          \and 
          G. Leung\inst{14}
          }

   \institute{Leibniz-Institut f\"ur Astrophysik Potsdam, An der Sternwarte 16, 14482 Potsdam, Germany\\
              \email{mkrumpe@aip.de}
   \and
      Max-Planck-Institut für Astronomie, K\"onigstuhl 17, D-69117 Heidelberg, Germany
   \and 
      Harvard-Smithsonian Center for Astrophysics, 60 Garden St., Cambridge, MA 02138, USA
   \and 
      Yale Center for Astronomy and Astrophysics, Yale University, 52 Hillhouse Ave, New Haven, CT 06511, USA
   \and 
      School of Physics \& Astronomy, Cardiff University, Queens Buildings, The Parade, Cardiff, CF24 3AA, UK
   \and 
      Gemini Observatory, Northern Operations Center, 670 N. A’ohoku Pl., Hilo, Hawaii, 96720, USA
   \and 
      Max-Planck-Institut für extraterrestische Physik, Giessenbachstr. 1, 85748 Garching, Germany
   \and 
      I. Physikalisches Institut, Universität zu K\"oln, Z\"ulpicher Straße 77, 50937 K\"oln, Germany
   \and 
      LERMA, Observatoire de Paris, College de France, PSL, CNRS, Sorbonne Univ., UPMC, 75014 Paris, France
   \and 
      Sydney Institute for Astronomy, School of Physics, University of Sydney, NSW 2006, Australia
   \and 
      ARC Centre of Excellence for All-sky Astrophysics (CAASTRO), Australia
   \and 
      Max-Planck-Institut für Radioastronomie, Auf dem H\"ugel 69, 53121 Bonn, Germany
   \and 
      Instituto de Astrofísica de Andaluc\'{i}a, Glorieta de las Astronom\'{i}a s/n, 18008 Granada, Spain
    \and 
      Center for Astrophysics and Space Sciences, University of California, San Diego, 9500 Gilman Dr., La Jolla, CA 92093, USA
     }

   \date{\textit{Draft as of} \today}
   
   \abstract{After changing optical AGN type from 1.9 to 1 in 1984, the AGN Mrk 1018 recently reverted back to its type 1.9 state. Our ongoing monitoring now reveals that the AGN has halted its dramatic dimming, reaching a minimum around October 2016. The minimum was followed by an outburst rising with $\sim$0.25 U-band mag/month. The rebrightening lasted at least until February 2017, as confirmed by joint \textit{Chandra} and \textit{Hubble} observations. Monitoring was resumed
in July 2017 after the source emerged from sunblock, at which point the AGN was found only $\sim$0.4 mag brighter than its minimum. The intermittent outburst was accompanied by the appearance of a red wing asymmetry in broad-line shape, indicative of an inhomogeneous broad-line region.
The current flickering brightness of Mrk 1018 following its rapid fading 
either suggests that the source has reignited, remains variable at a low level, 
or may continue dimming over the next few years. Distinguishing between 
these possibilities requires continuous multiwavelength monitoring.}


   \maketitle
   
\section{Introduction}
During a galaxy's lifetime, various physical processes, many of which remain poorly understood, can either trigger or shut down a mass flow onto the central supermassive black hole (SMBH). Theoretical models \citep[e.g.,][]{Hopkins:2008a} predict that an SMBH accretion reservoir can be consumed in $\sim$$10^{4-5}$ years, such that the AGN transitions to a ``normal'' galaxy as the central engine shuts down. The AGN phase is likely episodic with a wide range of accretion ratios from one duty cycle to the next \citep[e.g.,][]{Novak:2011}. In this scenario, AGN are better described as {\it \textup{events}}  than as a discrete class of objects.

The term changing-look (CL) AGN originally referred to objects converting optical spectral type from Type 1 to Type 2 or vice
versa. More recently, the definition was relaxed to AGN showing significant changes in the spectral features.
It is not clear whether these objects are the extremes in a distribution of AGN variability or caused by some major discrete event in or around the central engine.
A clumpy accretion disk, that is, variable gas input, would naturally explain the 
absence of constant or only slowly varying accretion. 
The corresponding timescales would be shorter than the rotation period in the accretion disk, which is on the order of $\sim30$ years or shorter (depending on mass and distance from the SMBH). Accretion disks can also be warped, precessing, or lopsided. 
In the past, CL AGN have been explained either by variable obscuration along the line of sight, a tidal disruption event \citep[TDE,][]{Rees:1988}, or simply a drastic change in the intrinsic accretion rate. Understanding CL AGN therefore offers an opportunity to improve our understanding of the AGN central engine and accretion process around SMBHs.

Only a few cases of the original definition of CL AGN are known, such as NGC~7603 \citep{Tohline:1976}, NGC~7582 \citep{Aretxaga:1999}, NGC~2617 \citep{Shappee:2014,Bon:2016}, Mrk~590 \citep{Denney:2014}, HE~1135$-$2304 \citep{Parker:2016}, and SDSS J015957.64$+$003310.5 \citep{LaMassa:2015, Merloni:2015}. Significant spectral changes are found for instance in 
SDSS J1011+5442 \citep{Runnoe:2016}, J01264--0839 and J2336+00172 \citep{Ruan:2016}, and iPTF 16bco \citep{Gezari:2017}. New monitoring projects now  actively hunt for more of these 
unique objects \citep{MacLeod:2016}.

The rarity of the changing-look behavior in AGN raises many questions: are these events simply rare and unique, or common but short
lived? 
Why and how does the AGN shut down, and how long does the shut-down process last? During shut-down, does the accretion disk disappear first or the X-ray corona that envelopes it? How fast can an AGN be reignited after it was shut down?

As part of the Close AGN Reference Survey (CARS; www.cars-survey.org), we have recently discovered a highly unique AGN in which to investigate these questions: \object{Mrk~1018}. The host galaxy of Mrk~1018 is an ongoing major merger and is one of the first known CL AGN, which transitioned from a Seyfert 1.9 to a Seyfert 1 nucleus around 1984 \citep{Cohen:1986}. 
The stellar mass of Mrk~1018 is log $M_*/M_{\odot} = 10.92$ \citep{Koss:2011} and its star formation 0.6 $M_{\odot}$ yr$^{-1}$ \citep{Shimizu:2017}. 
\cite{Bennert:2011} found log ($M_{\rm BH}/M_{\odot}) = 8.15$.
The current $L_{\rm bol}/L_{\rm edd}$ is around 0.03 \citep{McElroy:2016}. We note that because the broad-line region (BLR) of Mrk 1018 is currently not in an equilibrium or the virial factor has changed, recent $M_{\rm BH}$ estimates are significantly different (e.g., current faint phase: log ($M_{\rm BH}/M_{\odot}) = 7.4$; \citealt{McElroy:2016}).

\cite{McElroy:2016} and \cite{Husemann:2016b} reported that the nucleus significantly dimmed and changed type {\it back} to a Seyfert 1.9 between 2013--2015. 
Hence, Mrk 1018 is the first known CL AGN to undergo a full state change cycle, which we caught and monitored in the act of shutting down.  In this letter, we present recent optical, UV, and X-ray monitoring data of Mrk~1018 that 
reveal that the light curve has not continued dimming. Throughout, we adopt $H_0=70\,\mathrm{km\,s}^{-1}\,\mathrm{Mpc}^{-1}$, $\Omega_{\mathrm m} = 0.3$ and $\Omega_{\mathrm \Lambda} = 0.7$ cosmological parameters, and the AB magnitude system. Uncertainties represent 1$\sigma$ (68.3\%) confidence intervals unless otherwise stated.

\section{Optical monitoring}
\subsection{Dimming of Mrk 1018}
Shortly after the discovery of the optical dimming by a factor of $\sim$25 in mid-February 2016, follow-up observations with {\it Chandra} and {\it HST} were taken. The combined {\it Chandra} and archival {\it NuStar} spectra from February 2016 revealed that Mrk 1018 still shows no detectable absorption in the X-ray spectrum 
\citep{Husemann:2016b}, consistent with the bright Seyfert 1 phase (archival {\it Chandra} spectrum from 2010; ID: 12868, PI: Mushotzky). Hence, the current changing-look event cannot be caused by an obscuration event, but must instead be due to a significant drop in disk luminosity. \citet{LaMassa:2017} reanalyzed the {\it Chandra} data from 2010 and 2016 as well as the {\it NuStar} data. They confirmed that Mrk~1018 shows no line-of-sight X-ray absorption and concluded that the variation in the Fe K$\alpha$ equivalent width between 2010 and 2016 is due to a decreasing continuum flux and not an increase in the line flux. 

The significant drop in intrinsic disk luminosity is also confirmed by the decreasing luminosity at mid-IR wavelength as observed with {\it WISE} \citep{Sheng:2017}. 
The most recent NEOWISE data also indicate that the dimming stopped even in the mid-IR. However, no detailed quantification can be given because of the low spatial resolution and the substantial host galaxy contamination.
This suggests that the circumnuclear dust, usually referred to as the torus, is exposed to a significantly lower incident radiation field as well. Since the  brightness of Mrk~1018 appeared to be stable for several decades and the dimming does not follow an expected $t^{-5/3}$ law, a TDE is unlikely. Two possible scenarios to explain the significant luminosity variations of Mrk~1018 on a timescale of a few years were discussed in \citet{Husemann:2016b}: 1) a binary SMBH interaction, or 2) self-regulated accretion through accretion disk winds.

\subsection{Reaching a minimum in October 2016}
In response to our discovery that Mrk~1018 has faded by an order of magnitude within just a few years, we started an optical monitoring program with the VLT-VIMOS instrument \citep{LeFevre:2003}. Deep $U$ -band images with 600\,s total integration time each were obtained between June 2016 and January 2017 and from July 2017 until now with an average cadence of $\text{about two}$ weeks. The images were reduced with the ESO VIMOS pipeline and accurate relative photometric calibration was achieved with a reference star $60\arcsec$ away from the nucleus. 

\begin{figure}
 \includegraphics[width=\hsize]{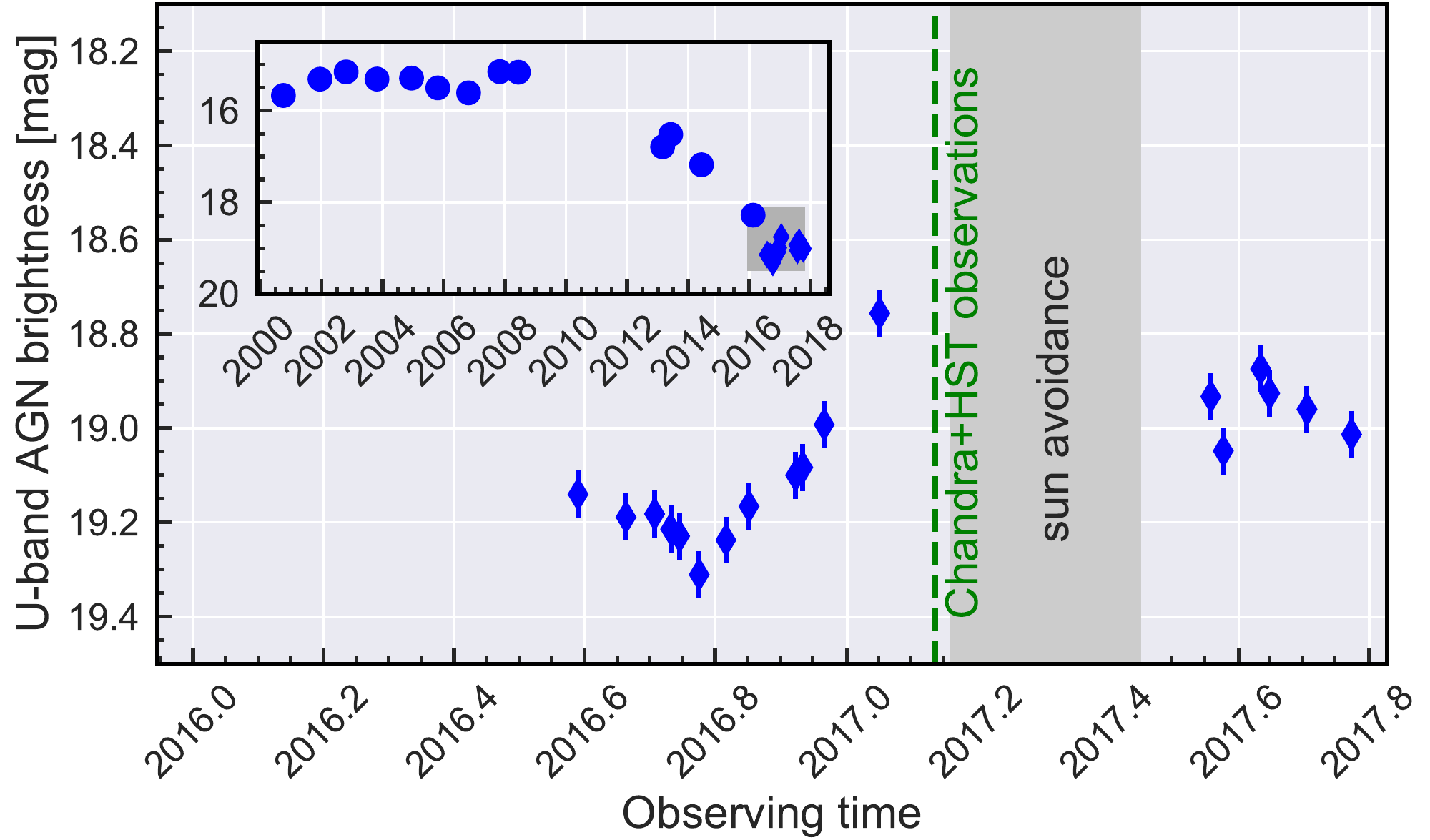}
 \caption{$U$ -band light curve of the AGN nucleus between June 2016 and October 2017 after subtraction of the host galaxy contribution. The inset shows the entire $U$ -band light curve since 2000 for comparison as presented in \citep{McElroy:2016}. The gray area in the inset highlights the coverage of the zoom-in axes. The 
 substantial sunblock period in 2017, where it is unobservable, is indicated by the black hatched area. The date of our joint \textit{Chandra} and \textit{HST} 
 observation is shown as the green dashed line.}
 \label{fig:Mrk1018_VIMOS}
\end{figure}

We performed a 2D imaging fitting with {\tt galfit} \citep{Peng:2010} of all images to decompose the light of the AGN nucleus and the host galaxy. After masking out the tidal features of this major merger system, the main body of the host galaxy can be reasonably well described by a single S\'ercic profile. We fixed the parameters of the model to $n=3.9$, $r_e=14.37''$, $b/a=0.85$ and PA=$-85^\circ$. These values are guided by a best-fit model of the deep coadded SDSS Stripe 82 model that we used previously to generate the long-term light curve in \citet{McElroy:2016}. The resulting $U$ -band light curve of the AGN component is shown in Fig.~\ref{fig:Mrk1018_VIMOS}. The $U$ -band magnitude of the host galaxy component, which is the only other free parameter left for the 2D fitting, is 16.09\,mag with an rms error of 0.01\,mag. We adopt a 1$\sigma$ error of 0.05\,mag for the AGN brightness to account for systematic uncertainties of the AGN-host galaxy deblending. 
A minimum brightness level of $m_{U,\mathrm{AGN}}\sim 19.3$\,mag was reached in October 2016. 

\subsection{Outburst of Mrk~1018}
Unexpectedly, the nucleus began to brighten again beginning in October 2016, by $\sim$0.25 mag/month. The rebrightening lasted at least until February 2017, when our joint \textit{Chandra} and \textit{HST} observations were executed (see Sections~\ref{X-ray} and \ref{Hubble}) just before this equatorial source passed into sunblock. Monitoring was resumed in July 2017, revealing a variable brightness of $\pm 0.1$ mag around $m_{U,\mathrm{AGN}} \sim 18.9$\,mag. The current brightness level is still $\sim$$0.4$\,mag higher than during the minimum state in October 2016 (Fig.~\ref{fig:Mrk1018_VIMOS}). We have confirmed that the host galaxy brightness remains constant within a nominal uncertainty.
The outburst was apparently very short with an unknown peak brightness. 
\begin{figure}
\includegraphics[width=0.5\textwidth]{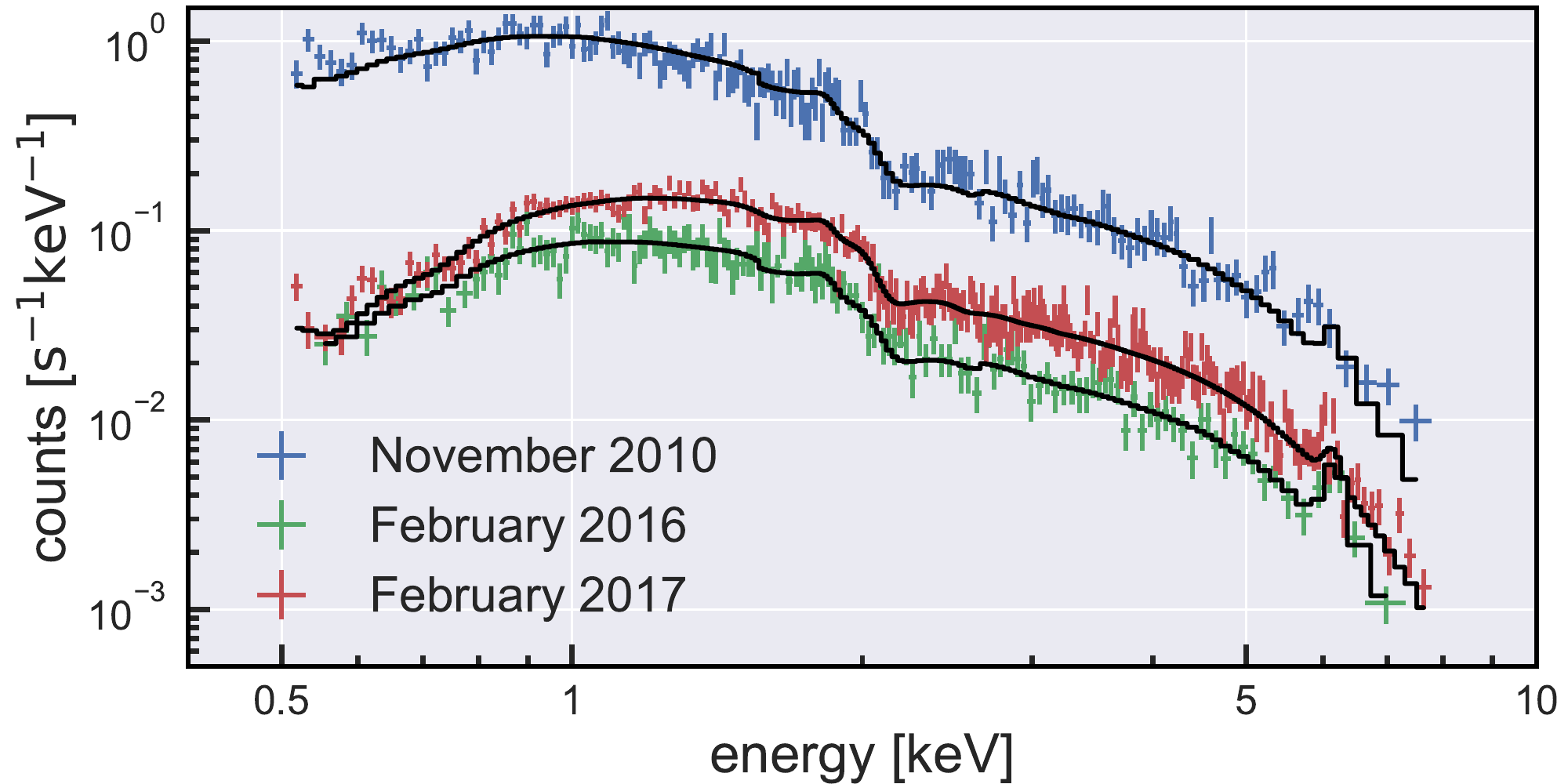}
\caption{Comparison of the X-ray spectra of Mrk 1018 at three epochs (2010, 2016, and 2017). For all data, we show the best-fit model. These observed spectra are not unfolded by the telescope response.}
\label{fig:xray}
\end{figure}
\begin{table}
\caption{Best-fit model parameters of \textit{Chandra} spectra.}
\label{table:xray}      
\centering                         
\begin{small}
\begin{tabular}{lccc}\hline\hline  
Epoch   &  $\Gamma$ &  $f_\mathrm{0.5-2keV}$ & $f_\mathrm{2-10keV}$ \\ 
&     & \multicolumn{2}{c}{[$10^{-12}\mathrm{erg\,cm}^{-2}\mathrm{s}^{-1}$]}\\
\hline                        
2010 Nov & 1.70$\pm0.03$ & 4.96$\pm0.15$  & 9.26$\pm0.19$ \\     
2016 Feb & 1.62$\pm0.03$ & 0.58$\pm0.02$  & 1.25$\pm0.03$ \\
2017 Feb & 1.63$\pm0.02$ & 1.12$\pm0.02$  & 2.32$\pm0.03$\\\hline                                   
\end{tabular}
\end{small}
\tablefoot{The intrinsic absorption ($N_H$) is consistent with zero at all epochs. Fluxes are corrected for Galactic absorption.}    
\end{table}

\section{X-ray view of Mrk 1018 during the outburst}
\label{X-ray}
The latest {\it Chandra} observation of Mrk~1018 was taken on 2017-02-17 (ID: 19560, PI: Krumpe) with ACIS-S for $\sim$50 ks. The spectra were extracted with the CIAO package (4.9) and the latest CALDB files (4.7.3) using standard setting for point sources. Since pile-up only affects the central pixel with a likelihood of $\sim$1\%, we ignored the effects of pile-up and created a standard source and background spectrum. The spectrum was grouped with a minimum binning of 20 counts. 

We fit the X-ray spectra using {\tt Xspec} version 12.9.1e \citep{Arnaud:1996} in the 0.5--8 keV energy range. We used the cosmic abundances of \citet{Wilms:2000} and the photoelectric absorption cross sections provided by \citet{Verner:1996}. We fit a model consisting of an intrinsically absorbed power law, a narrow Gaussian line profile fixed at a rest-frame energy of 6.4 keV (width $\sigma=0.1$ keV), and absorption by Galactic neutral hydrogen \citep[$N_{\rm HI,Gal} = 2.43 \times 10^{20}$ cm$^{-2}$,][]{Kalberla:2005}. The 90\% upper limit on the intrinsic $N_{\rm H}$ is $1 \times 10^{20}$ cm$^{-2}$ , consistent with no neutral absorption along the line of sight in excess over the Galactic column. We tentatively detect a narrow Fe K$\alpha$ line with 2.6$\sigma$ (EW=$0.14^{+0.05}_{-0.06}$ keV). The best-fit photon index (slope of the X-ray spectrum) is $\Gamma = 1.63\pm 0.02$ ($\Delta \chi^2/d.o.f.=298/271$, freezing the intrinsic $N_{\rm H}=0$). This model yields Galactic absorption-corrected fluxes of $f_\mathrm{0.5-2\,keV}=1.1 \times 10^{-12}$ erg cm$^{-2}$ s$^{-1}$ and $f_\mathrm{2-10\,keV}=2.3 \times 10^{-12}$ erg cm$^{-2}$ s$^{-1}$, which corresponds to rest-frame luminosities of $L_\mathrm{0.5-2\,keV}=4.6 \times 10^{42}$ erg s$^{-1}$ and $L_\mathrm{2-10\,keV}=9.5 \times 10^{42}$ erg s$^{-1}$.

We reanalyzed the 2010 and 2016 {\it Chandra} data in the same manner as described above to minimize systematic uncertainties. Best-fit model parameters for the three {\it Chandra} epochs are given in Table~\ref{table:xray}. A direct comparison of all Mrk~1018 data obtained with {\it Chandra} is shown in Fig.~\ref{fig:xray}. The photon indices in 2016 and 2017 are consistent with each other. Owing to the limited signal-to-noise ratio in all \textit{Chandra} observations, the decrease in photon index in 2016 and 2017 compared to the 2010 observation is not significant ($\sim$2$\sigma$) when considering the uncertainties. 

The data from 2017 can be equally well fit ($\Delta \chi^2/d.o.f.=298/272$) when using the above-mentioned model {\tt tbabs*(zpowerlaw+zgauss}) frozen to the 2016 best-fit model and only allowing the 2017 spectrum to vary globally in normalization, as well as allowing
the Fe line flux to vary freely. The best fit reveals a scaling factor of $1.92 \pm 0.02$. 
No obvious systematic difference between fitting the data with the 2016 best-fit model or a free fit are found. 
The fact that the broad-band X-ray spectrum only shows a scaling in flux but not in shape is also verified when comparing the change in 0.5--2 and 2--10 keV fluxes between the two observations.

\begin{figure}
 \includegraphics[width=\hsize]{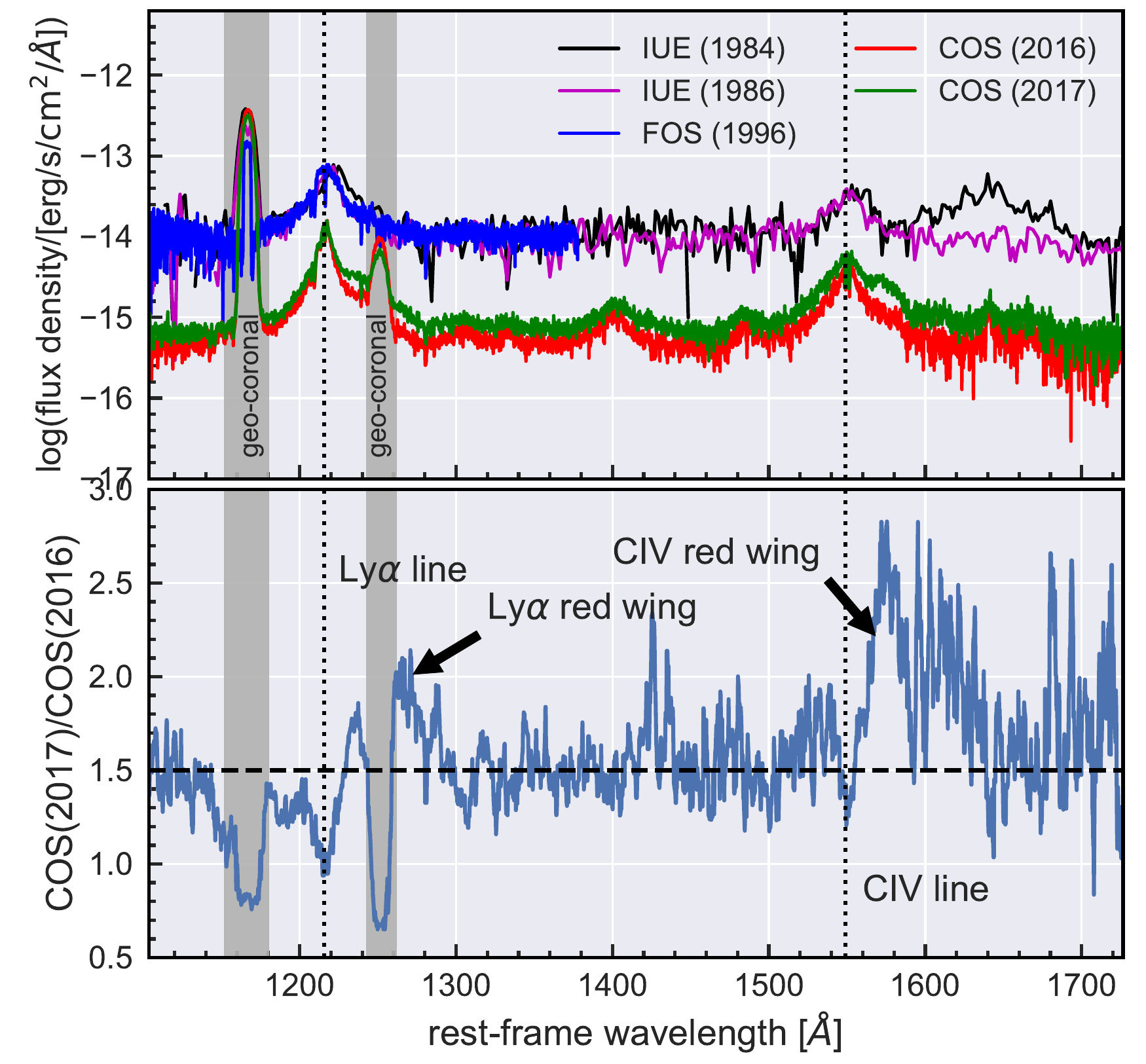}
 \caption{\textit{Top}: FUV spectra of Mrk~1018 from 1984 until 2017. \textit{Bottom}: Ratio of the 2017 and 2016 \textit{HST}/COS spectra, with a 357-day temporal baseline. Median filtering is applied to suppress the continuum noise. The horizontal dashed line highlights a continuum scale factor of 1.5.}
 \label{fig:Mrk1018_FUV}
\end{figure}
\begin{figure}
 \includegraphics[width=\hsize]{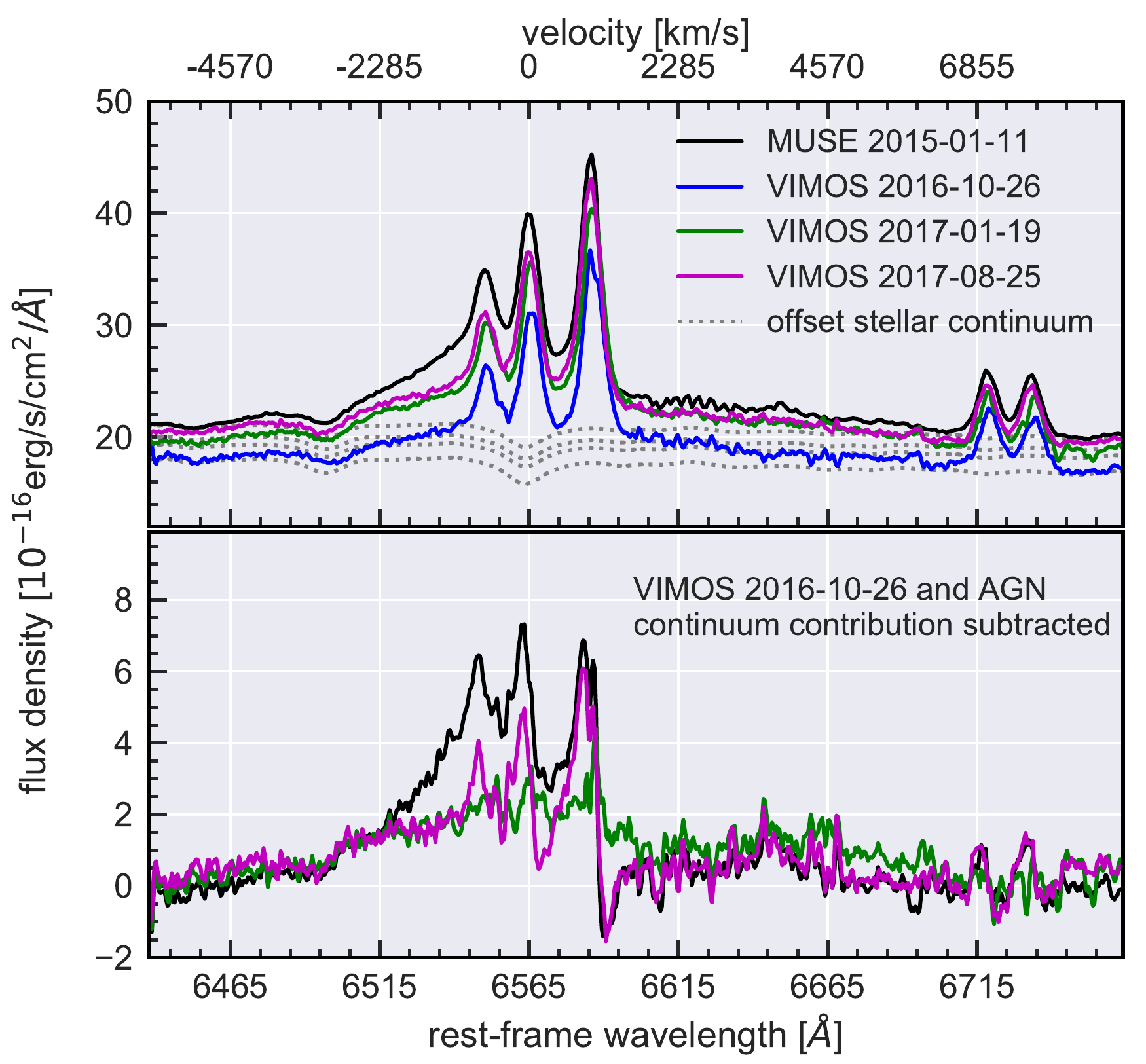}
 \caption{{\it Top:} Evolution of the broad H$\alpha$ line of Mrk~1018. Aperture spectra (3\arcsec\ diameter) from MUSE and three representative VIMOS IFU observations. Matching offset stellar continua model are shown. {\it Bottom:} Difference spectra with respect to the minimum spectrum in October 2016. A constant stellar continuum is subtracted in each spectrum, but offset in flux to take the varying AGN continuum emission
into account.}
 \label{fig:Mrk1018_Ha}
\end{figure}

\section{Broad emission-line variations}\label{Hubble}

Far-UV (FUV) spectroscopy with \textit{HST}/COS using the G140L grism was obtained quasi-simultaneously with the \textit{Chandra} observation on 2017 February 18 (ID: 14853, PI: M. Krumpe) over two\,orbits. The data were processed with the standard COS pipeline, and the calibrated spectrum is shown in Fig.~\ref{fig:Mrk1018_FUV} (upper panel) compared to several archival FUV spectra. The FUV continuum flux increased by a factor of $\sim$1.5 between 2016 and 2017 (357 days), which is similar to but slightly lower than the increase in X-ray luminosity during the outburst. Interestingly, the broad Ly$\alpha$ {\sl and} \ion{C}{IV} lines show prominent shape changes between 2016 and 2017. We highlight the changes in the lower panel of Fig.~\ref{fig:Mrk1018_FUV} by dividing the spectra. While the central Ly$\alpha$ line core dimmed, the red wing of the Ly$\alpha$ {\sl and} \ion{C}{IV} lines significantly increased in flux.

We have also undertaken spectroscopic monitoring of the broad H$\alpha$ line with VIMOS in IFU mode using the high-resolution red grism over a FoV of $27\arcsec\times27\arcsec$.  A reference star 60\arcsec\ away was observed twice together with Mrk~1018, enabling accurate relative spectrophotometry. The VIMOS IFU data were reduced with our own {\tt Py3D} IFU data reduction package \citep{Husemann:2013b, Husemann:2014}. In Fig.~\ref{fig:Mrk1018_Ha} we compare the spectrum observed with MUSE in 2015 with three representative spectra obtained around the minimum in 2016, the last one in January 2017 and the first spectrum after sun block in July 2017. In contrast to the FUV spectrum, which entirely scales with AGN luminosity, the optical spectra include additive components from the host galaxy and narrow emission-line regions. Thus, Fig.~\ref{fig:Mrk1018_FUV} uses the ratio between spectra and Fig.~\ref{fig:Mrk1018_Ha} the flux difference for comparison.

Despite some narrow-line residuals, the spectra reveal that the blue H$\alpha$ line asymmetry seen with MUSE in 2015 has faded away. This blue asymmetry was not present during the bright Type 1 AGN phase, and has likely appeared now because the \textit{\textup{red wing}} of H$\alpha$ disappeared amid the dimming phase, as discussed in \citet{McElroy:2016}. More importantly, we confirm that the prominent red wing asymmetry in Ly$\alpha$ and \ion{C}{IV} is not present in H$\alpha$ line 
only one month before the {\it HST} observations were taken. 
This confirms that the extreme red wing asymmetry in the FUV lines is directly associated with the outburst. It corresponds either to an asymmetry in the BLR with a distance of less than four light weeks from the SMBH, or it might correspond to a distinct accreting object with its own BLR.

\section{Discussion and summary}
Unexpectedly, the nucleus of Mrk~1018 reached a minimum brightness in October 2016 followed by a short outburst.  The nucleus is currently $\Delta U \sim0.4$ mag brighter than during the minimum a few months ago, which indicates that Mrk~1018 interrupted or ended its extraordinary dimming phase. If the global dimming really has ended, the minimum brightness of Mrk~1018 was only an order of magnitude fainter than in the bright phase. This would be a substantially different behavior than observed for CL AGN Mrk~590 \citep{Denney:2014}, which dimmed by a factor of $\sim$100 in a few years after a three-decade-long bright Type 1 AGN phase. This raises the question as to whether the same physical processes are causing the rapid disk luminosity drop in the CL AGN Mrk~1018 and Mrk~590.  

The evolution in the shape of the broad emission lines encodes crucial information on the structure of the BLR around the AGN. Many AGN display significant changes in their broad-line shapes, including CL AGN, for example, NGC~5548 \citep{Bon:2016} or NGC~2617 \citep{Oknyansky:2017}. In the case of Mrk~1018, the blue wing asymmetry in the broad-line shape during the global dimming phase \citep{McElroy:2016} faded away after 2015, 
whereas a prominent red wing asymmetry appeared less than four weeks before the brief outburst. 
These line-shape variations may be related to asymmetries in the BLR distribution, to accretion disk winds, or even to a close binary AGN that varies independently in accretion disk luminosity and BLR response. 

Between the bright Seyfert 1 phase and 2016, the UV and X-ray flux dropped by a factor of $\sim$17 and $\sim$8, respectively (\citealt{Husemann:2016b}), but these observations were taken several years apart in the bright phase. 
The quasi-simultaneous UV and X-ray observations from 2016 and 2017 showed that the flux increases during the apparent outburst in 2017 shared similar amplitudes (factors of 1.5 and 1.9, respectively). Thus, the X-ray corona and accretion disk are responding in a nearly identical fashion. Both regions must be within a light travel time of shorter than or around a year, and perhaps as short as a few months,  given the short outburst time as seen in the $U$ band. 

The details of the underlying physical processes driving the apparently complex evolution of Mrk~1018  cannot be constrained with the currently available data. A better understanding of the central engine of Mrk 1018 now awaits 
reverberation mapping of the BLR over a far longer temporal baseline, combined with multiwavelength monitoring of the accretion disk emission and its interaction with the X-ray corona. That the dimming has possibly halted enables continuous monitoring of the AGN with a high signal-to-noise ratio, at least for the time being. The accretion disk of Mrk~1018 is nevertheless likely in an unstable state, and may brighten or fade with little warning. 
We therefore expect that ongoing radio through X-ray monitoring of Mrk~1018 will yield exciting discoveries over the coming years.

\begin{acknowledgements}
MK acknowledges support from DLR grant 50OR1802 and DFG grant KR 3338/3-1.
GRT acknowledges support from NASA through the Einstein Postdoctoral Fellowship Award Number PF-150128, issued by the Chandra X-ray Observatory Center, which is operated by the Smithsonian Astrophysical Observatory for and on behalf of NASA under contract NAS8-03060. MAPT acknowledges support from the Spanish MINECO through grants AYA2012-38491-C02-02 and AYA2015-63939-C2-1-P. TAD acknowledges support from a Science and Technology Facilities Council Ernest Rutherford Fellowship. Parts of this research were conducted by the Australian Research Council Centre of Excellence for All-sky Astrophysics (CAASTRO), through project number CE110001020. GL acknowledges support  provided by the National Aeronautics and Space Administration through Chandra Award Number G07-18111X issued by the Chandra X-ray Observatory Center (NAS8-03060).
We thank Alison Coil for helpful discussions and suggestions. 
\end{acknowledgements}

\bibliographystyle{aa}
\bibliography{references}

\end{document}